\documentclass[reprint,superscriptaddress,aps,prl,amsmath,amssymb,nofootinbib]{revtex4-2}
\usepackage{natbib}
\usepackage{orcidlink}
\usepackage{graphicx}
\usepackage{physics}
\usepackage{xcolor}
\usepackage{xspace}
\usepackage{siunitx}
\usepackage{hyperref}
\graphicspath{{./figs/}}

\newcommand{\neff}{N_\textrm{eff}}
\newcommand{\omgw}{\Omega_\textrm{GW}}
\newcommand{\omgwo}{\Omega_\textrm{GW,0}}
\newcommand{\rot}{rule of thumb\xspace}

\usepackage{aas_macros}

\usepackage{xcolor}

\begin{document}

\bibliographystyle{apsrev4-1}

\title{CMB and energy conservation limits on nanohertz gravitational waves}

\author{David Wright\,\orcidlink{0000-0003-1562-4679}}
\affiliation{Department of Physics, Oregon State University, Corvallis, OR 97331, USA}

\author{John T. Giblin, Jr\,\orcidlink{0000-0003-1505-8670}}
\affiliation{Department of Physics, Kenyon College, Gambier, OH 43022 USA}
\affiliation{Department of Physics/CERCA/Institute for the Science of Origins, Case Western Reserve University, Cleveland, OH 44106, USA}
\affiliation{Center for Cosmology and AstroParticle Physics (CCAPP) and Department of Physics, Ohio State University, Columbus, OH 43210, USA}

\author{Jeffrey Hazboun\,\orcidlink{0000-0003-2742-3321}}
\affiliation{Department of Physics, Oregon State University, Corvallis, OR 97331, USA}

\begin{abstract}
  The recent evidence for a stochastic gravitational wave background (GWB) in the nanohertz band, announced by pulsar timing array (PTA) collaborations around the world, has been posited to be sourced by either a population of supermassive black holes binaries or perturbations of spacetime near the inflationary era, generated by a zoo of various new physical phenomena.
  Gravitational waves (GWs) from these latter models would be explained by extensions to the standard model of cosmology and possibly to the standard model of particle physics.
  While PTA datasets can be used to characterize the parameter spaces of these models, energy conservation and limits from the cosmic microwave background (CMB) can be used \emph{a priori} to bound those parameter spaces.
  Here we demonstrate that taking a simple rule for energy conservation and using CMB bounds on the radiation energy density can set stringent limits on the parameters for these models.
\end{abstract}

\maketitle

{\sl Introduction:} Pulsar timing arrays \cite{foster+1990} have recently seen the first evidence for a stochastic background of gravitational waves (GWs) in the nanohertz band \cite{InternationalPulsarTimingArray:2023mzf,NANOGrav:2023gor,EPTA:2023fyk,Reardon:2023gzh}. The main indicator of these GWs is the telltale Helling--Downs correlations \cite{hd83} between pulsar timing datasets dependent on the angular separation of the pulsars on the sky, as seen from Earth. The source of these GWs is currently unknown \cite{NANOGrav:2023gor,NANOGrav:2023hvm,NANOGrav:2023hfp}; Neither the spectral characterization or the searches for single-sources or anisotropy \cite{NANOGrav:2023tcn,NANOGrav:2023pdq} are sensitive enough in current datasets to say whether the GWs are sourced by a population of astrophysical supermassive black hole binaries (SMBHBs) or a cosmological process.

Stochastic backgrounds are predicted from a number of cosmological scenarios. These include from preheating at the end of inflation \cite{Khlebnikov:1997di, Easther:2006gt, Easther:2006vd, Garcia-Bellido:2007nns, Easther:2007vj, Dufaux:2007pt, Dufaux:2008dn, Dufaux:2010cf, Adshead:2018doq, Adshead:2019lbr, Adshead:2019igv, ElBourakadi:2022lqf, Cosme:2022htl,Mansfield:2023sqp,Adshead:2024ykw,Brummer:2024ejc} scalar-induced gravitational waves \cite{Domenech:2021ztg,Yuan:2021qgz,Garcia:2023eol,Altavista:2023zhw,Afzal:2024xci,Perna:2024ehx,Caprini:2024gyk,Zhou:2024doz},  bubble collisions \cite{Kosowsky:1991ua,Kosowsky:1992vn,Child:2012qg} or phase transitions \cite{Kosowsky:1992rz,Kamionkowski:1993fg,Caprini:2006jb,Caprini:2015zlo,Caprini:2019egz,Hindmarsh:2020hop}.
These early universe GWs would have contributed to \(\Omega_{r}\) at recombination and should have observable imprints in CMB observables.
The allowable excess energy in CMB observations of \(\Omega_{r}\) is often parameterized by \(\Delta\neff\), the effective number of neutrino species beyond the standard model (SM) expectation.
Thus, \(\neff\) constraints from CMB experiments \citep{Planck:2018vyg} offer a way to bound the integrated energy density of an early universe GWB.

Going beyond total energy density bounds, \cite{Giblin:2014gra} presents a method of estimating the maximum spectral energy density in gravitational waves, starting from conservation of energy considerations; this ``rule of thumb'' assesses the feasibility of a cosmological GWB based on a minimal-assumptions parameterization. The \rot takes into account properties of the source that \(\neff\) bounds do not, such as how quadrupolar the energy density is.
While the \rot is not a \emph{bound} it offers information complementary to \( \neff\) constraints.


{\sl Models and Analysis:} In this letter, we demonstrate that existing arguments, bounds from the CMB and the \rot, provide important insight into the feasibilty that current (and future) PTA data might be cosmological in origin.  To demonstrate the power of this argument,  we report on two fiducial, representative models: GWs produced by sound waves in a first-order phase transition (FOPT) and scalar-induced GWs (SIGWs) from the injection of a Dirac delta function into the scalar power spectrum.  Our results with other models can be found on GitHub\footnote{\url{https://github.com/davecwright3/rule-of-thumb}}.

To do this analysis, we very closely follow the same model parameterization and code implementations as the NANOGrav 15yr new physics analyses \citep{ng15_models,Mitridate:2023oar}; although the analysis can be repeated using any parameterization and should be extended to other models as they arise.

We will begin by reviewing two representative models, chosen from among the complete catalog in \cite{NANOGrav:2023gor}, outlining the models used therein to obtain best-fit parameters and posterior PDFs.  We will then describe the two proposed tests, the \rot and \(\neff\) bounds, before applying these tests to our two models.


{\sl Representative model---First-order phase transitions:}
The Standard Model (SM) predicts that cosmological phase transitions were not first order, but are smooth crossovers.
Therefore, the detection of a cosmological first order phase transition would present evidence for beyond standard model (BSM) physics.

First order phase transitions can produce gravitational waves from true-vacuum bubble collisions, plasma sound waves, and plasma turbulence \citep{Weir:2017wfa}.
The relative strength of each source is strongly model-dependent, so here we choose a source irrespective of a specific BSM theory where the dominant GW source in the FOPT comes from sound waves in a coupled fluid.

The GWB for such a FOPT is parameterized by the strength of the transition, \(\alpha_{*}\),  the average bubble separation at percolation in units of Hubble radius, \(R_{*}H_{*}\),  the percolation temperature, \(T_{*}\), and spectral shape parameters \(a,b,\) and \(c\).
Following the notation and parameterization of \textcite{NANOGrav:2023hvm}, the GWB PSD is
\begin{equation}
\Omega_{\textrm{FOPT}, 0}(f) = D \bar{\Omega}_{s}\Upsilon(\tau_\text{sw}) \left(\frac{\kappa_s \alpha_*}{1 + \alpha_*}\right)^2 (H_* R_*) S(f/f_s),
\end{equation}
where the spectral shape is assumed to be a \textit{broken~powerlaw}
\begin{equation}
S(x) = \frac{1}{N} \frac{\qty(a + b)^c}{\qty(bx^{-a/c} + ax^{b/c})^c}
\end{equation}
with normalization
\begin{align}
  N = \qty(\frac{b}{a})^{a/n} &\qty(\frac{nc}{b}\vphantom{\frac{b}{a}})^{c} \frac{\Gamma(a/n)\Gamma(b/n)}{n \Gamma(c)},\\
n &= (a+b)/c,
\end{align}
where \(\Gamma\) is the gamma function.
The efficiency of GW production is captured by \citep{Hindmarsh:2017gnf,Espinosa:2010hh}
\begin{align}
\bar{\Omega}_{s} &\approx 0.036 \label{eq:eff-omega}
\end{align}
and
\begin{align}
\kappa_s &= \frac{\alpha_{*}}{0.73 + 0.083\sqrt{\alpha_{*}} + \alpha_{*}} \label{eq:eff-kappa}.
\end{align}
The decrease of the GWB energy density due to redshifting is parameterized by
\begin{equation}
D = \frac{\pi^2T_{0}^{4}}{90 M_P^2H_0^2} \qty(\frac{g_{*,s}^{eq}}{g_{*,s}})^{4/3} \approx \num{1.67e-5}.
\end{equation}
Since GWs from sound waves are only emitted for a finite time,
we use the factor \(\Upsilon\) to apply a cutoff to the GWB PSD,
\begin{equation}
\Upsilon(\tau_{sw}) = 1 - \qty(1 + 2\tau_{sw} H_*)^{-1/2},
\end{equation}
where \(\tau_{sw}\) is the shock formation timescale
\begin{equation}
\tau_{sw} \approx \frac{4R_{*}\qty(1+\alpha_{*})}{3 \kappa_{s} \alpha_{*}}.
\end{equation}
We expect the PSD to peak at a frequency today \citep{Hindmarsh:2017gnf},
\begin{equation}
f_{s} \approx 48.5 \,\text{nHz} \, g_*^{1/2} \qty(\frac{g_{*,s}^{eq}}{g_{*,s}})^{1/3} \qty(\frac{T_*}{1\,\text{GeV}}) \frac{f_s^*}{H_*},
\end{equation}
where \(f_s^* = 1.58 / R_{*}\) is the peak frequency at time of emission \citep{Hindmarsh:2017gnf}.


{\sl Representative model---Scalar-induced gravitational waves:}
To first order, tensor and scalar perturbations are not coupled.
However, at second order large scalar perturbations can induce tensor perturbations and give rise to so-called scalar-induced gravitational waves.
These large scalar perturbations can take a myriad of forms depending on the inflationary theory proposed.
To remain agnostic to the microphysics while capturing the general behavior of SIGWs, the scalar perturbations are typically modeled as spectral features superposed on the scale-invariant power spectrum predicted by single-field slow-roll inflation.
Here, we choose to inject a Dirac delta function feature into the scalar power spectrum at a scale \(k_{\delta}\),
\begin{equation}
  \label{eq:delta-psd}
P_{R}\qty(k) = A\, \delta \qty(\ln{\frac{k}{k_{\delta}}}).
\end{equation}
The GWB PSDs at the time of emission, \(\Omega_{\textrm{SIGW}}(f)\), and today, \(\Omega_{\textrm{SIGW}, 0}(f) \), are 
\begin{equation}
  \label{eq:sigw-gwb-emission}
\Omega_{\textrm{SIGW}}(f) = \int_{0}^{\infty}\dd{v} \int_{\abs{1-v}}^{1+v} \dd{u} K\qty(u,v)P_{R}(uk)P_{R}(vk)
\end{equation}
and
\begin{equation}
  \label{eq:sigw-gwb-today}
  \Omega_{\textrm{SIGW}, 0}(f) = \Omega_{r} \qty(\frac{g_{*}(f)}{g_{*}^{0}})\qty(\frac{g_{*,s}^{0}}{g_{*,s}(f)})^{4/3}\Omega_{\textrm{SIGW}}(f),
\end{equation}
where the integration kernel \(K\qty(u,v)\) is \citep{Kohri:2018awv}
\begin{widetext}
\begin{eqnarray}
K(u, v) = \frac{3(4v^2 - (1 + v^2 - u^2)^2)^2(u^2 + v^2 - 3)^4}{1024 u^4v^8} \qty[\qty(\ln \abs{\frac{3 - (u + v)^2}{3 - (u - v)^2}} - \frac{4uv}{u^2 + v^2 - 3})^2+ \pi^2\Theta(u + v - \sqrt{3})].
\end{eqnarray}
\end{widetext}
The factors \(g_{*}\) and \(g_{*,s}\) account for the relativistic degrees of freedom in energy and entropy density, resp.
We numerically integrate Eq. \eqref{eq:sigw-gwb-today} for the frequencies considered in \citet{NANOGrav:2023gor}.


{\sl \(\neff\) Bounds:} The integrated energy bounds are derived from the CMB \(\neff\) limit.
We use the Planck 2018 \citep{Planck:2018vyg} TT, TE, EE+lowE+lensing+BAO measurement and set \(\Delta \neff=0.284\) from the reported 95\% limits on \(N_{\textrm{eff}}\).
The upper limit on \(\Omega_\textrm{GW}\) is then
\begin{equation}
\label{eq:integrated-energy-neff}
\int \frac{\dd f}{f}\ h^2 \Omega_\textrm{GW}\qty(f) < 5.6 \times 10^{-6} \Delta \neff,
\end{equation}
where we set \(h=0.674\).


{\sl Rule of Thumb:}
The \rot bounds are not truly ``bounds'' in the same manner as \(\neff\).
Instead, they represent the expected peak energy density given a specific parameterization of an early-universe GW source.

Following \citet{Giblin:2014gra}, we assume that our source is isotropic and has a characteristic scale, $k_*$, and width, $\sigma$.  Such a source can be approximated by a Gaussian in momentum space\footnote{The spectral shape changes only slightly when alternative parameterizations of the source are chosen. For example, models with broad spectral features may be more accurately described by a distribution other than a Gaussian, but the changes to the rule of thumb bounds are not substantial.} and has a stress energy tensor
\begin{equation}
  \label{eq:stress-energy}
  \widetilde{T}_{ij}\qty(\vec{k}) \approx \widetilde{T}\qty(\vec{k}) = A \exp \qty[-\frac{\qty(\abs{\vec{k}} - k_{*})^{2}}{2\sigma^{2}} ].
\end{equation}

From this, we can model the PSD of such a source today as
\begin{equation}
\label{eq:rot}
\omgwo(k_{*}) \approx 2.3 \times 10^{-4} \alpha^2\beta\omega^2 \frac{k_{*}}{\sigma} \qty(\frac{H}{k_{*}})^2,
\end{equation}
where \(k_{*}\) is the characteristic scale of the source, \(\alpha\) is the fraction of the closure density in the source at the time of emission, \(\beta\) is the magnitude of the anisotropic part of the source's stress-energy tensor, \(\omega\) is the equation of state parameter of the Universe at the time of the process, and \(\sigma\) is the standard deviation of the source's stress-energy tensor components \footnote{A full derivation is found in \cite{Giblin:2014gra}}.

As in \citep{Giblin:2014gra}, we set
\begin{equation}
\label{eq:rot-vals}
  \frac{k_{*}}{\sigma} = 2,\quad \frac{H}{k_{*}} = 0.01,
\end{equation}
which are motivated\footnote{The numerical values of these ratios should be chosen based on the models being considered (e.g. horizon-scale sources). Here we choose a reasonable upper-bound to demonstrate the method.} by reasonable expectations of phase transitions and resonant processes.
We use these values to arrive at a simplified expression for our \rot,
\begin{equation}
\label{eq:rot-final}
\omgwo(k_{*}) \approx 4.7 \times 10^{-8} \alpha^2\beta\omega^{2}.
\end{equation}
It is convenient, then, to use Eq.~\ref{eq:rot-final} to define pessimistic, realistic and optimistic values of $\Omega_{\rm GW}(k)$ given reasonable choices for $\alpha$, $\beta$ and $w$, as can be seen in Table~\ref{tab:rot-regions}.

{\sl Implementation:} For each model of interest, we use \textsc{SciPy}'s \citep{Virtanen:2019joe} orthogonal Latin hypercube (LH) routines to sample the parameter space.
We sample both the prior ranges and 68\% credible intervals (CI) from \citet{NANOGrav:2023hvm}.
Our integration and peak-finding routines\footnote{\url{https://github.com/davecwright3/rule-of-thumb}} utilize the \textsc{JAX} \citep{jax2018github} Python library for GPU acceleration and automatic vectorization over the LH samples.

In the \rot analysis, we calculate the peak energy density of the model for each LH sample over the frequency\footnote{The UV cutoff for horizon-scale sources is sensitive to the reheating temperature and number of e-folds during reheating. Here we assume that \(\omgw(f)\) extends to interferometer frequencies or decays quickly above its peak scale.} range \qtyrange{e-12}{e2}{\Hz}.
For the 68\% CI, we find the peak energy density of the maximum \textit{a posteriori} probability (MAP) values from \cite{NANOGrav:2023hvm}.
We also find the minimum and maximum peak energy density values within the 68\% CI.
The regions we consider for the \rot are given in Table \ref{tab:rot-regions}.

The \(\neff\) analysis largely follows the \rot analysis.
We again LH sample from the same regions, but we now integrate over frequency with Eq. \eqref{eq:integrated-energy-neff} using the same frequency bounds as the \rot analysis.
\begin{table}
  \caption{Regions for \rot bounds.}
  \label{tab:rot-regions}
\begin{tabular}{lcccc}
\hline
scenario & \(\alpha\)  & \(\beta\)& \(\omega\) &  \(\omgw(k_{*})\)  \\
\hline
optimistic  & 1  & 0.1  & 1/3 & \( 4.97\times10^{-10} \) \\
realistic  & 0.1& 0.03 & 1/3  & \( 1.49\times10^{-12} \) \\
pessimistic & 0.02& 0.005 & 1/3 & \( 9.93\times10^{-15} \) \\
\hline
\end{tabular}
\end{table}


\begin{figure}
  \includegraphics[width=\columnwidth]{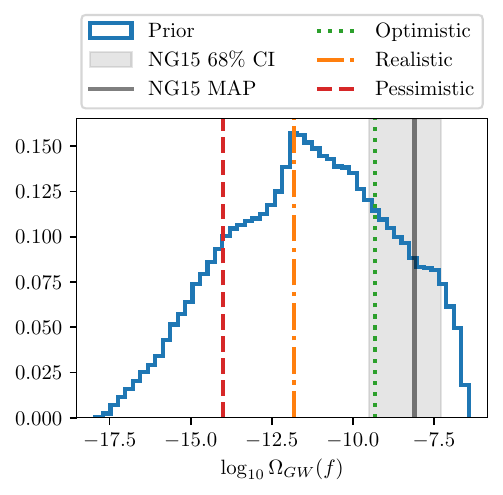}
  \caption{Rule of thumb regions for the FOPT model.
    The pessimistic, realistic, and optimistic choices of the parameters \(\alpha, \beta, \omega\) are defined in Table \ref{tab:rot-regions}.
    We take the MAP values and 68\% CI for this model from Table 4 of \citet{NANOGrav:2023hvm}.
  }
  \label{fig:pt-peak}
\end{figure}

\begin{figure}
  \includegraphics[width=\columnwidth]{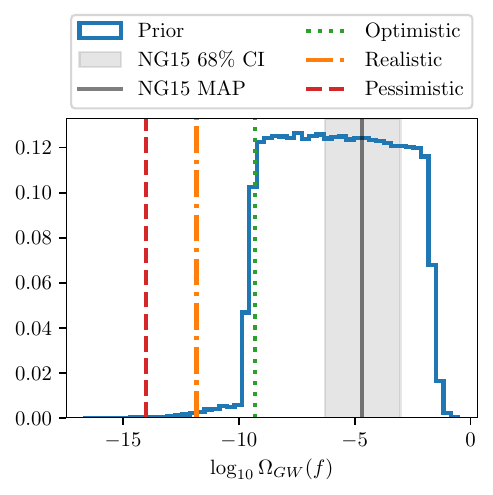}
  \caption{Rule of thumb regions for the SIGW model.
    The pessimistic, realistic, and optimistic choices of the parameters \(\alpha, \beta, \omega\) are defined in Table \ref{tab:rot-regions}.
    We take the MAP values and 68\% CI for this model from Table 4 of \citet{NANOGrav:2023hvm}.
  }
  \label{fig:sigw-peak}
\end{figure}

\begin{figure}
  \includegraphics[width=\columnwidth]{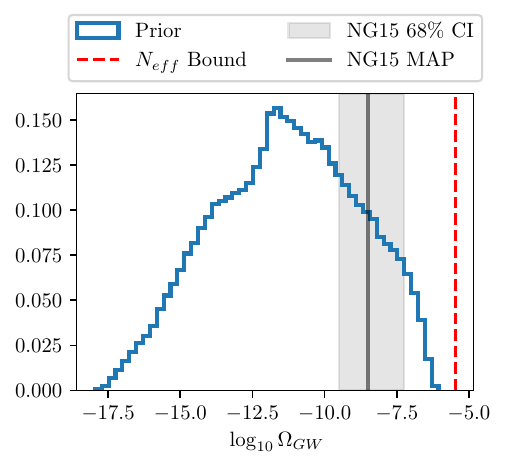}
  \caption{\(\neff\) bounds for the FOPT model using \citet{Planck:2018vyg} \(\neff\) constraints.
    We take the MAP values and 68\% CI for this model from Table 4 of \citet{NANOGrav:2023hvm}.
  }
  \label{fig:pt-int}
\end{figure}

\begin{figure}
  \includegraphics[width=\columnwidth]{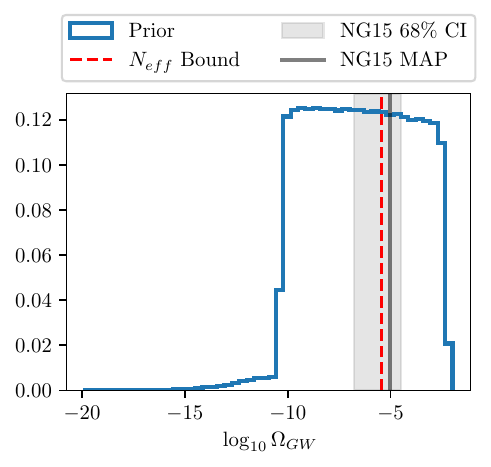}
  \caption{\(\neff\) bounds for the SIGW model using \citet{Planck:2018vyg} \(\neff\) constraints.
    We take the MAP values and 68\% CI for this model from Table 4 of \citet{NANOGrav:2023hvm}.
  }
  \label{fig:sigw-int}
\end{figure}
{\sl Results \& Discussion:} A stochastic background of gravitational waves can be a generic result of many cosmological processes---and the recent detection of a likely stochastic gravitational wave background provides the first-ever opportunity for these theoretical predictions in data.  Using spectral density models provided from theoretical predictions has led to systematic and thorough analysis of this data and provided statistics on the parameters of these models {\sl if} they are to explain the PTA data.

In this work, we ask whether or not these best-fit models are consistent with theoretical predictions by considering (1) whether the total integrated GW energy is allowed by measurements of the CMB and (2) if the estimated spectral density is consistent with expectations from estimates that consider conservation of energy at the time the GW are emitted\footnote{Of course, there are stochastic backgrounds that are not constrained by energy considerations, notably, inflationary gravitational waves---however, all pre-CMB gravitational waves are subject to \(\neff\) can constraints. Modified expansion histories are often invoked to enhance scalar perturbations and give rise to observable SIGWs or primordial black holes (PBH).}.

We examine two cases in detail.  For the FOPT, the MAP parameter values from \citet{NANOGrav:2023hvm}, as well as most of the parameters' 68\% CIs, are consistent with current limits on  \(\neff\) from the CMB.  However, the MAP values, as well as a significant portion of the parameters' 68\% CIs, lie above the ``optimistic'' limit from the \rot.  As such, if a FOPT is responsible for the PTA data, it would need to violate the assumptions of that maximum energy estimate.

The SIGW model, on the other hand, has MAP parameter values \citep{NANOGrav:2023hvm} that produce a background currently ruled-out by \(\neff\) bounds, as well as 68\% CIs for the model parameters that generate backgrounds almost entirely above the ``optimistic'' limit from the \rot.  Therefore, we conclude, that the PTA data are inconsistent with a GW background from this model.

Going forward, we argue that it is imperative for analyses of PTA data to take seriously these considerations when comparing PTA data (or any data measuring stochastic gravitational wave backgrounds) to theoretical models.  In particular, increasingly accurate measurements of \(\neff\) provide a direct, orthogonal test as to whether these gravitational waves were present at recombination---a requirement for cosmological gravitational waves produced at these early times.
Even if not ruled out directly by \(\neff\), data analysis should also consider whether these best-fit models are expected from energy considerations, e.g. the \rot. Evading these limits requires increasingly exotic physics, such as stiff equations of state or modified expansion histories.  For generic scenarios, models that exceed ``optimistic" limits should address which assumptions are violated.

\bibliography{references.bib}
\end{document}